\documentclass[11pt]{article}
\usepackage{amsmath, amssymb, amscd, amsthm, amsfonts}
\usepackage{graphicx}
\usepackage{enumerate}
\usepackage{hyperref}
\usepackage{natbib}
\usepackage{amssymb}
\usepackage{xr}
\usepackage{threeparttablex}
\usepackage{graphicx,amssymb,booktabs,multirow}
\usepackage{graphicx,amssymb,booktabs,multirow}
\usepackage{url} 

\makeatletter
\newcommand*{\addFileDependency}[1]{
	\typeout{(#1)}
	\@addtofilelist{#1}
	\IfFileExists{#1}{}{\typeout{No file #1.}}
}
\makeatother

\makeatletter
\newcommand*{\indep}{%
	\mathbin{%
		\mathpalette{\@indep}{}%
	}%
}
\newcommand*{\nindep}{%
	\mathbin{
		\mathpalette{\@indep}{\not}
	}%
}
\newcommand*{\@indep}[2]{%
	\sbox0{$#1\perp\m@th$}
	\sbox2{$#1=$}
	\sbox4{$#1\vcenter{}$}
	\rlap{\copy0}
	\dimen@=\dimexpr\ht2-\ht4-.2pt\relax
	\kern\dimen@
	{#2}%
	\kern\dimen@
	\copy0 
}

\newtheorem{theorem}{Theorem}
\newtheorem{lemma}{Lemma}
\newtheorem{remark}{Remark}

\newtheorem{assumption}{Assumption}

\newtheorem{example}{Example}

\def\cov{\textnormal{Cov}}
\def\var{\textnormal{Var}}

\newenvironment{sequation*}{\begin{equation*}\small}{\end{equation*}}

\newenvironment{tequation*}{\begin{equation*}\tiny}{\end{equation*}}

\usepackage{mathrsfs}

\def\ols{\rm{ols}}

\usepackage{accents}

\def\T{{ \mathrm{\scriptscriptstyle T} }}

\def\eif{\textnormal{eff}}
\def\calM{\mathcal{M}}

\newcommand\Pn{{\mathbb{P}_n}}
\newcommand\mr{\textnormal{mr}}
\newcommand\union{\textnormal{union}}
\newcommand\unif{\textnormal{Unif}}

\newcommand\cf{\textnormal{cf}}

\newcommand\calS{\mathcal{S}}

\graphicspath{ {./figures/} }
\usepackage{color}

\def\ho{{\rm h}}
\def\eq{{\rm e}}

\allowdisplaybreaks

\title{\bf Identification and multiply robust estimation of causal effects  via instrumental variables from an auxiliary  population}
		\author{Wei Li and Jiapeng Liu\hspace{.4cm}\\
			Center for Applied Statistics and School of Statistics, Renmin University of China \hspace{.4cm}\\
			and \\
			Peng Ding \\
			Department of Statistics, University of California, Berkeley\hspace{.4cm}\\
			and\\
			Zhi Geng\\
			School of Mathematics and Statistics, Beijing Technology and Business University
		}
		\date{}

\begin{document}
		
	
	{
		
	\maketitle
	\begin{abstract}

Estimating causal effects in a target population  with unmeasured
confounders is challenging, especially when  instrumental variables (IVs) are unavailable. However, IVs from auxiliary populations with similar problems can
help infer causal effects in the target population. While the homogeneous
conditional average treatment effect  assumption has been widely used for
effect transportability, it has not been explored in IV-based data fusion. We include it as a basic approach, though it may be biased when treatment effect heterogeneity exists. As an alternative approach, we introduce the equi-confounding assumption  that the unmeasured confounding bias remains the same after adjusting for
observed covariates, while allowing conditional average treatment effects to differ across populations. This allows us to identify the confounding bias in the auxiliary population and remove it from the treatment-outcome association in the target population to recover the causal effect. We develop multiply robust estimators under both approaches
and demonstrate them
through simulation studies and a real data application.

	\end{abstract}
	
	\noindent%

	\newpage
	\section{Introduction}
	\label{sec:intro}
	It is challenging to estimate causal effect of a treatment on an outcome in observational studies with unmeasured confounding. A commonly employed strategy to address unmeasured confounding is to use instrumental variables (IVs). However,
	IVs may be unavailable in the   target  population. In such cases, auxiliary studies with complete IV information can be leveraged to estimate causal effects in the target population.
For example, 
	researchers are interested in the effect of education on earnings in a specific industry within a developing country where  no suitable IVs are available in the observed data.  
	Similar studies  in developed countries have used changes in compulsory schooling laws as IVs to study the same causal problem \citep{angrist1991does,oreopoulos2006estimating}.  
	By appropriately using  studies from the developed countries, we can  infer the causal effect of education on earnings in the developing country  without  IVs.

	Integrating data from other sources has become increasingly popular for achieving valid causal inference in  the target population  \citep{
			bareinboim2016causal,
		yang2019combining,
			hu2022paradoxes,
        li2024calibrated,colnet2024causal}. 
	A strand of literature 
	focuses on generalizing or transporting findings from auxiliary randomized trials to the target population based on various versions of the selection exchangeability assumption \citep{stuart2011use,
        pearl2014external,
		buchanan2018generalizing,
		dahabreh2020extending,
		yang2023elastic}.
        This assumption typically requires that the conditional average treatment effects (CATEs) given observed covariates are equal across both  populations. 
	In the absence of randomized trials, some studies  use  an auxiliary dataset with  supplemental confounding information to mitigate
	estimation bias in the target population \citep{
		yang2019combining,luo2025multiply}. In particular, \cite{yang2019combining} proposed a general framework to estimate
	causal effects in the setting where the target data have unmeasured
	confounders, but the smaller  auxiliary data provide
	supplementary information on these confounders. 
	
    Another line of research leverages auxiliary data with IVs to establish
	identification  and estimation of causal effects  when the target data  lack some important variables for subsequent analysis \citep{
		ridder2007econometrics,
		zhao2019two,shuai2023identifying}. Specifically, in the classical two-sample IV framework, the target data contain information on the outcome, IV,  and covariates, but lack information on the treatment;  the auxiliary data provide information on the treatment, IV, and covariates, but lack information on the outcome. This method
	has been widely used in econometrics and social sciences, and  mendelian randomization studies with genetic factors as IVs \citep{
		graham2016efficient,
		shu2020improved,zhao2019two}. Recently, \cite{sun2022semiparametric} developed semiparametric efficient estimation of average treatment effects in the two-sample IV framework.
	However, this framework requires measurements of IVs in both the target and auxiliary data, which differs from our setting where only the auxiliary data contain IVs.  

    In this paper, we address the problem of identifying and estimating causal effects in a target population, where unmeasured confounding may be present but IVs are unavailable. To tackle this challenge, we leverage auxiliary data from a population where valid IVs exist. Our approach is built on a set of assumptions that ensure the identifiability of the CATE in the auxiliary population using IVs. In particular, we adopt the  no-additive-interaction assumption between unmeasured confounders and treatment \citep{wang2018bounded}, which ensures the identification of the average causal effects with IVs. {While \citet{wang2018bounded} studied IV-based identification in a single  dataset, our work addresses a more complex data integration setting, where IVs are only available in the auxiliary population. From this perspective, our framework can be viewed as their data integration counterpart.}
    To extend the identification from the auxiliary population to the target population, additional assumptions are required. 
    
    We propose two distinct transportability approaches for identification. The homogeneous CATE assumption is canonical for transporting results from randomized trials to observational studies but has not been  explored in IV-based data fusion problems. We include it as the basic approach. While  it offers a simple way to transport effects, it may be biased when treatment
effect heterogeneity exists.
    As an alternative approach, we introduce a novel equi-confounding assumption, which allows for heterogeneous CATEs while ensuring that after adjusting for observed covariates, the unmeasured confounding bias remains the same across populations. 
    Unlike the homogeneous CATE assumption, which directly sets identical conditional
treatment effects across populations, equi-confounding leverages additional information by explicitly estimating both the unmeasured confounding bias and the treatment-outcome association. Specifically, the unmeasured confounding bias can be computed in the auxiliary
population and subsequently removed from the observed treatment-outcome association in the target population to identify the average causal effect. 
This  adjustment
enables effect transportability while allowing for treatment effect heterogeneity.
 Under each identifiability approach,  we derive the efficient influence function to motivate the corresponding semiparametric  estimator. These estimators are multiply robust in that they remain consistent even if the observed data model is partially misspecified. We also show that they are consistent and asymptotically normal, even when machine learning techniques are used for nuisance models estimation. 
    
	
	The rest of this paper is organized as follows. Section \ref{sec:identification} introduces the notation, assumptions, and identification results. Section \ref{sec:mr} presents the semiparametric  multiply robust estimators and  their machine learning analogues.
	Section \ref{sec:simulation} uses simulation studies to evaluate the finite-sample performance of the proposed estimators. Section \ref{sec:app} applies our approaches to estimate the average causal effect of  smoking on physical functional status in the higher-income individuals.  We conclude  in Section \ref{sec:discussion} and relegate proofs to the supplementary material. 

	\section{Notation, assumptions, and identification}
	\label{sec:identification}
	Consider $n$ individuals merged from two different sources. Let $R$ denote the source indicator, where $R=0$ if a subject is from the  target population, and $R=1$ if from the auxiliary  population. In the target data, we observe a binary treatment $X$ with 0 and 1 for the labels of control and active treatments, respectively. Let $Y$ denote an outcome of interest and $V$ denote a vector of pre-treatment covariates. We assume that $U$ contains all unmeasured common causes of $X$ and $Y$.
	We use the potential outcomes
	framework to define causal effects.
	Let $Y_x$ denote the potential outcome if a subject, possibly contrary to the fact, were assigned to treatment
	$x$. We impose the consistency assumption that the observed outcome is $Y=XY_1+(1-X)Y_0$.
	We are interested in the average treatment effect of the target population: 
	\begin{align*}
		\tau=E(Y_1-Y_0\mid R=0).
	\end{align*}
    
	In the auxiliary sample, besides the treatment $X$, the outcome $Y$, and  observed covariates $V$, a binary IV $Z$ is also observed. {Let $O=(X,Y,Z,V,R)$ if $R=1$, and $O=(X,Y,V,R)$ if $R=0$ denote the observed data.}
    For $r=0,1$, define $\beta_r(V)$ to be the CATEs in the two populations: 
    \begin{align*}
        \beta_r(V)=E(Y_1-Y_0\mid V, R=r).
    \end{align*}
    We use $f(\cdot)$ to denote the probability density or mass function of a random variable (vector). Table \ref{tab:notation} summarizes some key notation used throughout the paper.


   \begin{table}[t!]
		\caption{Some key notation used throughout the paper. \\
		}
		\label{tab:notation}
		\centering
		\resizebox{0.98\textwidth}{!}{
			\begin{tabular}{ccc}
				\toprule
				Notation      &    Definition    & Description     \\ 
				\hline
	$\omega(V)$ & $f(R=1\mid V)$ & Sampling mechanism\vspace{-2mm}\\
    $\gamma(V)$ & $\{1-\omega(V)\}/\omega(V)$ & Odds function \vspace{-2mm}\\
    $q$ & $f(R=0)$ or $1-E\{\omega(V)\}$  & Target data proportion\vspace{-2mm}\\
	$\mu_R(V)$ & $E(X\mid V,R)$ & Treatment propensity score\vspace{-2mm}\\
    $\eta(X,V,R)$ or $\eta$ & $X-\mu_R(V)$ & Treatment regression residual\vspace{-2mm}\\
    $\sigma_R^2(V)$ &$\var(X\mid V,R)$
    & Conditional variance of treatment\vspace{-2mm}\\
     $\beta_R(V)$ &$E(Y_1-Y_0\mid V,R)$
    & Conditional average treatment effect\vspace{-2mm}\\
    $\pi(V)$ & $f(Z=1\mid V,R=1)$ & Conditional probability of $Z=1$  in auxiliary data\vspace{-2mm}\\
    $\mu_0^X(V)$ & $E(X\mid Z=0,V,R=1)$ & Conditional treatment mean given $(Z=0,V)$ in auxiliary data \vspace{-2mm}\\
    $\mu_{0}^Y(V)$ &$E(Y\mid Z=0,V,R=1)$ & Conditional outcome mean given $(Z=0,V)$ in auxiliary data\vspace{-2mm}\\
       $\delta^X(V)$ &$E(X\mid Z=1,V,R=1)-E(X\mid Z=0,V,R=1)$ & Conditional treatment mean difference in auxiliary data\vspace{-2mm}\\
   $\delta^Y(V)$ &$E(Y\mid Z=1,V,R=1)-E(Y\mid Z=0,V,R=1)$ & Conditional outcome mean difference in auxiliary data\vspace{-2mm}\\
   $ \phi_R(V)$ &$E\{Y-\beta_R(V)X\mid V,R\}$ & Conditional mean function of the residual $Y-\beta_R(V)X$\vspace{-2mm}\\
   $\rho(V)$&$E[\eta \{Y-\beta_R(V)X\}\mid V,R=0]$ & Conditional mean function of $\eta \{Y-\beta_R(V)X\}$ in target data\\
				\bottomrule
			\end{tabular}
		}
	\end{table}

   \subsection{Preliminary assumptions}\label{subsec:preliminary}

   In this subsection, we  introduce several commonly-used assumptions that ensure the identification of 
$\beta_1(V)$ using IVs.

	\begin{assumption}[Latent ignorability and overlap]\label{ass:latent}
		(i) $Y_x\indep X\mid V,U,R$ for $x=0,1$; (ii) $f(X=x\mid V,U,R)>0$ for $x=0,1$.
	\end{assumption}
Assumption~\ref{ass:latent}(i) means that the treatment $X$ is latent ignorable given observed covariates $V$ and unmeasured variables $U$ in both the target and auxiliary populations. 
Assumption~\ref{ass:latent}(ii) ensures sufficient overlap between the covariate distributions of the treatment and control groups. {In addition, it also implies overlap in treatment assignment between the target and auxiliary populations.}
    To identify \( \beta_1(V) \), Assumption~\ref{ass:latent} is required only for individuals with \( R=1 \). We also include latent ignorability and overlap for \( R=0 \) in Assumption~\ref{ass:latent}, as it is required by the  identification approach
    introduced in Section \ref{subsec:equi-iden} later.  
    
	\begin{assumption}[IV for auxiliary population]\label{ass:iv}
		(i) $Z\indep U\mid V,R=1$; (ii) $Z\indep Y\mid X,V,U,R=1$; (iii) $Z\nindep X\mid V, R=1$.
	\end{assumption}

	Assumption~\ref{ass:iv} characterizes standard conditions for a valid IV $Z$ in the auxiliary population. Assumption~\ref{ass:iv}(i) requires the IV  to be independent of unmeasured confounders $U$ given $V$. Assumption~\ref{ass:iv}(ii) formalizes no direct effect of the IV  on the outcome $Y$.  Assumption~\ref{ass:iv}(iii) ensures that the IV  is correlated with the treatment $X$ even after conditioning on $V$. Although these assumptions are generally untestable without additional restrictions, possible IVs may be obtained through some natural or quasi-experiment in observational studies \citep{	  
    angrist1991does,
    angrist1996identification,
baiocchi2014instrumental}.

	Assumptions \ref{ass:latent} and \ref{ass:iv} are commonly made in the causal inference literature. However, they are insufficient to identify $\beta_1(V)$ in the auxiliary population. To identify $\beta_1(V)$, we introduce the following no-interaction assumption  \citep{wang2018bounded}:
    \begin{assumption}[No  additive $U$-$X$ interaction] \label{ass:no-interact}
    $E(Y_1-Y_0\mid V,U,R)=\beta_R(V)$. 
\end{assumption}

The identification of $\beta_1(V)$ requires Assumption \ref{ass:no-interact} to hold only  for $R=1$.
Similar to Assumption \ref{ass:latent},  
we also impose Assumption \ref{ass:no-interact} for \( R=0 \) to support later discussion.
Assumption~\ref{ass:no-interact} allows unmeasured confounders to affect the control potential outcome but assumes they do not  modify the treatment effect on the outcome. It is reasonable when key effect modifiers are observed and adjusted for, reducing the role of unmeasured confounders in modifying the treatment effect.
 Similar no-interaction conditions are widely used in the literature for identifying the average treatment effect with valid or invalid IVs \citep{wang2018bounded,sun2022semiparametric,
 tchetgen2021genius,sun2022selective}. A key implication of Assumption~\ref{ass:no-interact} is the need to collect as many predictors of the treatment and outcome as possible to ensure that no residual effect modification involving 
$U$ remains within strata of the observed covariates $V$.



Assumptions \ref{ass:latent} and \ref{ass:no-interact} with $R=1$, along with the IV Assumption \ref{ass:iv}, are sufficient to identify $\beta_1(V)$ as 
     \begin{align}
		\label{eqn:beta1}
		\beta_1(V)= \frac{E(Y\mid Z=1,V,R=1)-E(Y\mid Z=0,V,R=1)}{E(X\mid Z=1,V,R=1)-E(X\mid Z=0,V,R=1)}.
	\end{align}
    The nonparametric representation in \eqref{eqn:beta1} corresponds to the  conditional Wald estimand \citep{wang2018bounded, sun2022semiparametric}.

\begin{assumption}[Selection positivity]\label{ass:positive}
    $f(R=r\mid V)>0$ for $r=0,1$.
\end{assumption}
Assumption \ref{ass:positive} requires  overlap of the observed covariate
distribution between the target and auxiliary 
populations. Similar positivity assumptions have been widely used in the data fusion literature \citep{
dahabreh2020extending,
degtiar2023review,
shi2023data,colnet2024causal}.

 \subsection{Identification under homogeneous CATE}\label{subsec:homo-iden}
   By definition, the causal parameter is given by \( \tau = E\{\beta_0(V) \mid R = 0\} \). To identify \( \tau \), additional assumptions are required to establish a connection between \( \beta_0(V) \) and \( \beta_1(V) \).  
A straightforward approach involves imposing selection exchangeability or weaker mean exchangeability assumptions, which assumes that the target and auxiliary populations are homogeneous when conditioned on the observed covariates \( V \). 

\begin{assumption}[Homogeneous CATE]\label{ass:homo}  
    \( \beta_0(V) = \beta_1(V) \).  
\end{assumption}  

Similar assumptions have been widely adopted in the literature for generalizing findings from randomized controlled trials to the target population. For example, \cite{degtiar2023review}, \cite{shi2023data}, and \cite{colnet2024causal} incorporated these assumptions in their frameworks. Here, we adapt Assumption \ref{ass:homo} to the  data fusion setting involving IVs. It requires  covariates
to capture all treatment effect modifiers that are shifted between the two populations. It would be
satisfied if the auxiliary sample is randomly drawn from the target population or if all shifting effect modifiers are measured.




    \begin{lemma}
    \label{lem:homo}
        Suppose that Assumptions \ref{ass:latent} and \ref{ass:no-interact} for $R=1$, Assumptions \ref{ass:iv}, \ref{ass:positive}, and \ref{ass:homo} hold. Then  $\tau$ is identified {by $\tau = \tau^h$, where} 
        \begin{align}\label{eqn:identify-homo}
{\tau^\ho = E\{\beta_1(V)\mid R=0\},}
        \end{align}
       {and the superscript ``h"  indicates the homogeneous CATE assumption.
       }
    \end{lemma}

\subsection{Identification under equi-confounding}\label{subsec:equi-iden}

The homogeneous CATE assumption has been widely used in existing data fusion methods, but it may be violated when the auxiliary data come from heterogeneous populations.
Below we introduce a novel identification approach that allows for different  CATEs between the target and auxiliary populations. 
    
\begin{assumption}[Equi-confounding]
\label{ass:equiconf}
    $\cov(X,Y_0\mid V,R=0)=\cov(X,Y_0\mid V,R=1).$
\end{assumption}

The covariance $\cov(X,Y_0\mid V,R)$ in Assumption~\ref{ass:equiconf} measures unmeasured confounding bias. Notably, when there is no unmeasured confounding, it is  zero.
Assumption~\ref{ass:equiconf} posits that, conditional on observed covariates, the magnitude of unmeasured confounding bias is the same in  the target and auxiliary populations. {Under Assumptions \ref{ass:latent} and \ref{ass:no-interact}, we show in Lemma S2 of the supplementary material that  
$\cov(X, Y_1 - Y_0 \mid V, R)= 0$,  
which implies  \(\cov(X, Y_0 \mid V, R) = \cov(X, Y_1 \mid V, R)\) for both populations. Therefore, requiring the equality of the covariance for \(Y_0\)  in Assumption \ref{ass:equiconf} is  equivalent to requiring it for \(Y_1\).
}

Assumption~\ref{ass:equiconf} would be testable if IV data were available in the target population. However, we focus on a more challenging and realistic setting where IV data are only observed in the auxiliary population. In our setting, this assumption is not testable.
Assumption \ref{ass:equiconf} is reasonable when the key unmeasured confounders operate through similar mechanisms in both populations. This is more likely if the confounders are stable characteristics such as genetic traits, rather than context-dependent factors. Additionally, when rich observed covariates account for major differences between two populations, the remaining confounding bias is less likely to vary significantly.


 The following example provides some  sufficient conditions for Assumptions \ref{ass:no-interact} and \ref{ass:equiconf}.
Let $\varepsilon(V,U)=E(X\mid V,U,R=1)-E(X\mid V,U,R=0)$ denote the difference of the treatment propensity scores given all confounders $(V,U)$ beween the two populations.
	

	\begin{example}\label{exam:model}
		Suppose  $U\indep R\mid V$ and the following structural equation  model holds:
		\begin{align}
			E(Y\mid X,V,U,R)=&~\zeta_0(V,U)+\zeta(V)R+\beta_R(V)X,
			\label{eqn:modelforY}
		\end{align}
		where  $\cov\{\zeta_0(V,U),\varepsilon(V,U)\mid V,R\}=0$,
		and
		$\{\zeta_0(\cdot),\zeta(\cdot)\}$ are unknown scalar functions. Then under Assumption \ref{ass:latent}, Assumptions \ref{ass:no-interact} and \ref{ass:equiconf} are satisfied. 

           Since the variables $(X,R)$ are both binary, model~\eqref{eqn:modelforY} is equivalent to
$ E(Y\mid X,V,U,R)=\zeta_0(V,U)+g(X,V,R)$
    for an arbitrary function $g(\cdot)$.  
    Model \eqref{eqn:modelforY} essentially
	requires that there are no additive $U$-$R$ and $U$-$X$ interactions. A sufficient condition for the zero-covariance constraint under model \eqref{eqn:modelforY} is $\varepsilon(V,U)=\varepsilon(V)$, which implies no additive $U$-$R$ interaction in predicting the treatment  conditional on observed covariates. As discussed earlier, similar no-interaction conditions  appeared in \cite{wang2018bounded} and \cite{tchetgen2021genius}.
	\end{example}

    Now we present the identification formula of $\tau$ under the equi-confounding assumption.
	\begin{theorem}\label{thm:identification}
		Suppose that Assumptions~\ref{ass:latent}--\ref{ass:no-interact}, \ref{ass:positive}, and \ref{ass:equiconf} hold. Then $\beta_0(V)$ is identified as
		\begin{align}\label{eqn:beta0}
			\beta_0(V)=E\bigg\{ \frac{\beta_1(V)\sigma^2_1(V)}{\sigma_0^2(V)}-\frac{(2R-1)\eta Y}{f(R\mid V)\sigma_0^2(V)}\mid V\bigg\},
		\end{align}
       {where all notation is defined in Table~\ref{tab:notation}. In addition,
        $\tau$ is identified  by $\tau = \tau^e$, where} 
		\begin{align}\label{eqn:identify-equi}
			{\tau^\eq = E\Bigg[\frac{(1-R)\beta_1(V)\sigma^2_1(V)}{q\sigma^2_0(V)}-\bigg\{\frac{R}{f(R\mid V)}-1 \bigg\}\frac{\eta Y}{q\sigma^2_0(V)} \Bigg],}
		\end{align}	
        {and the superscript ``e"  indicates the equi-confounding assumption.
        }
	\end{theorem}

     Theorem~\ref{thm:identification} provides a novel  
	 identification result about $\beta_0(V)$  under equi-confounding.   It provides  one identification formula for $\tau$. There are  alternative identification formulas based on different components of the observed data likelihood, which are omitted due to space constraints.
     
     The IV Assumption~\ref{ass:iv} is specifically imposed to identify the CATE  in the auxiliary population $\beta_1(V)$, which can be replaced by any alternative conditions as long as $\beta_1(V)$ is guaranteed to be identifiable. For example, if the unmeasured confounders $U$ are fully observable in the auxiliary population, as considered  in \citet{yang2019combining}, then under Assumption~\ref{ass:latent},  $\beta_1(V)$ can be identified as
	\begin{align*}
		\beta_1(V)=E\big\{E(Y\mid X=1,V,U,R=1)-E(Y\mid X=0,V,U,R=1)\mid V,R=1\big\}.
	\end{align*}
    Similarly, when negative control variables are available in the auxiliary population, $\beta_1(V)$ can  be also identifiable under certain conditions \citep{Miao2018Identifying,tchetgen2020introduction}.
	In these scenarios, the identification expressions of $\tau$  in Lemma \ref{lem:homo} and Theorem~\ref{thm:identification} remain unchanged except that the expression of $\beta_1(V)$ therein is updated. 
     
     In proving Theorem \ref{thm:identification}, we  express $\beta_0(V)$ as:
\begin{align}\label{eqn:did-beta0}
    \beta_0(V) = \beta_0^{\ols}(V) - \big\{\beta_1^{\ols}(V) - \beta_1(V)\big\}
    \frac{\sigma^2_1(V)}{\sigma_0^2(V)},
\end{align}
where $\beta_r^{\ols}(V)$  denotes the ordinary least squares (OLS) estimand for the coefficient of $X$ obtained by the simple regression of $Y$ on $X$, stratified by $V$ and $R=r$, i.e., $\beta_r^{\ols}(V)=\sigma_r^{-2}(V)\cov(X,Y\mid V,R=r)$ for $r=0,1$.
     If covariates $V$ are omitted and 
     the variances of the
     treatment are equal across  populations such that $\sigma_1^2 = \sigma_0^2$, equation \eqref{eqn:did-beta0} simplifies to: 
\begin{align}
\label{eqn:simple-beta0}
\beta_0 = \beta_0^{\ols} - (\beta_1^{\ols} - \beta_1).
\end{align}
This simplification offers a more intuitive interpretation of the identification result in Theorem~\ref{thm:identification}. 
Specifically, the auxiliary data enable us to express the unmeasured confounding bias using observable quantities (i.e., $\beta_1^{\ols} - \beta_1$), which is then subtracted from the identified
treatment-outcome association $\beta_0^{\ols}$ in the target population to yield an expression for the causal effect. This  transportability strategy relies on the equi-confounding Assumption \ref{ass:equiconf}.
Because Assumption \ref{ass:equiconf} parallels the additive equi-confounding bias assumptions used in identification approaches based on difference-in-differences or negative outcome controls \citep{card1994minimum,angrist2009mostly,lipsitch2010negative,tchetgen2014control,sofer2016negative}, the identification formula in \eqref{eqn:simple-beta0} naturally resembles a difference-in-differences adjustment.

\begin{remark}\label{remark:equi}
In general, the variance ratio term \( \sigma_1^2(V)/\sigma_0^2(V) \) in the identification formula~\eqref{eqn:did-beta0} cannot be omitted under our current equi-confounding Assumption~\ref{ass:equiconf}, as doing so would lead to bias. To eliminate this ratio, one might consider replacing Assumption~\ref{ass:equiconf} with the following alternative:
\begin{align}\label{eqn:sel-confounding-nostandard}
	\beta_0^{\ols}(V) - \beta_0(V) = \beta_1^{\ols}(V) - \beta_1(V).
\end{align}
Assumption \eqref{eqn:sel-confounding-nostandard} may appear reasonable and ensures a simpler identification formula. However, 
it is difficult to find interpretable conditions under which the two sides are  equal. In particular, when the homogeneous CATE assumption holds, \eqref{eqn:sel-confounding-nostandard} further requires that $\beta_0^{\ols}(V) = \beta_1^{\ols}(V)$.
For example, suppose that Assumption \ref{ass:latent} holds, and consider the following  model:
\begin{equation}
\begin{aligned}\label{eqn:linear-sem}
    R,V,U~\mathop\sim^{iid}&~\textnormal{Bernoulli}(p),\\
    E(Y\mid X,V,U,R)=&~\zeta_0+\zeta_1 R+\zeta_2 V+\zeta_3U+\beta X,\\
    E(X\mid Z,V,U,R)=&~\iota_0+\iota_1 R+\iota_2 V+\iota_3U+\iota_4 RZ.
\end{aligned}
\end{equation}
Since  the IV $Z$ is only observed  when $R=1$, the model for $X$ includes the interaction term $RZ$. We assume that 
$R,V,U$ follow the same Bernoulli distribution for simplicity.
Under $\eqref{eqn:linear-sem}$, we have $\beta_0(V)=\beta_1(V)=\beta$. Then  \eqref{eqn:sel-confounding-nostandard} requires $\beta_0^{\ols}(V) = \beta_1^{\ols}(V)$.
In fact, we have $\beta_r^{\ols}(V)=p(1-p)\zeta_3\iota_3 \sigma^2_r(V)$ under model \eqref{eqn:linear-sem}. Therefore, \eqref{eqn:sel-confounding-nostandard} does not hold if $\zeta_3\iota_3\neq 0$ and $\sigma_1^2(V)\neq \sigma_0^2(V)$. The inequality in variance occurs when $\iota_1\neq 0$ or $\iota_4\neq 0$. 
In contrast, Assumption~\ref{ass:equiconf} is satisfied under \eqref{eqn:linear-sem}. More generally, Assumption \ref{ass:equiconf}
is implied by the conditions in Example~\ref{exam:model}, whereas \eqref{eqn:sel-confounding-nostandard} typically does not hold in that setting. Details are given in Section S3.4 of the supplementary material.
\end{remark}

 Recently, \cite{li2025data} proposed a method to incorporate weakly aligned  data sources that do not satisfy selection exchangeability or the weaker homogeneous CATE assumptions. Their approach, based on selection bias models,  requires at least one known auxiliary data source to properly align with the target population, ensuring the identification of the target parameter. Their  goal is to improve statistical efficiency by incorporating additional weakly aligned data sources. In contrast, our equi-confounding approach
 does not require the auxiliary data source to be pre-aligned with the target population. Instead, our approach establishes a novel identification result based on the assumption that the unmeasured confounding bias remains stable across populations. {This paper also bears some connection to \cite{li2024efficient}, particularly Example 3 in their work. While \cite{li2024efficient} assumed that the confounding bias lies in a smooth, finite-dimensional parametric class, we allow the bias to be an arbitrary and possibly nonparametric
function.}

\subsection{Comparison}
\label{subsec:comparison}

The identification strategies in Lemma~\ref{lem:homo} and Theorem~\ref{thm:identification} offer two distinct approaches for transporting causal effects from the auxiliary population to the target population. The homogeneous CATE assumption directly equates the conditional treatment effects across populations, i.e., $\beta_0(V) = \beta_1(V)$. In contrast, the equi-confounding assumption posits that  unmeasured confounding biases are the same in the two populations, conditional on observed covariates. This assumption does not require the CATEs  to be equal across populations.
As shown in~\eqref{eqn:simple-beta0} for the simplified setting where covariates are omitted and  the variances of the
     treatment are the same in the two  populations, the equi-confounding approach incorporates a difference-in-differences type correction. It leverages additional information from OLS estimates and explicitly accounts for confounding bias using data from the auxiliary population, thereby enabling effect transportability even when treatment effect heterogeneity exists.

 The choice between these two approaches depends on the study context. The homogeneous CATE assumption works well when treatment effect heterogeneity is small or explained by observed covariates.  The equi-confounding assumption is more appropriate when unmeasured confounders operate similarly in both populations, even if the CATEs differ. Since these assumptions are untestable and not mutually exclusive, they may both hold in certain settings.
If domain knowledge strongly supports one assumption, the corresponding approach should be preferred. Otherwise, researchers should carefully consider which assumption aligns best with the data and study design.
 When possible, applying both approaches and comparing results, as demonstrated in our real data analysis, can provide additional robustness and strengthen conclusions.

Since both approaches rely on key transportability assumptions, we can  conduct sensitivity analyses to evaluate their robustness. To assess the impact of potential violation of the homogeneous CATE Assumption \ref{ass:homo}, we can introduce the 
sensitivity parameter as the difference of the two CATEs across populations
and modify the identification formula in Lemma~\ref{lem:homo} accordingly  to conduct sensitivity analysis. 
To
 assess the impact of potential violations of equi-confounding Assumption~\ref{ass:equiconf}, we introduce the following sensitivity parameter: $
\xi(V) = \text{Cov}(X, Y_0 \mid V, R=0) - \text{Cov}(X, Y_0 \mid V, R=1)$,
which captures the difference in unmeasured confounding bias between the target and auxiliary populations. 
Under Assumption~\ref{ass:equiconf}, we have $\xi(V) = 0$, and hence, a large absolute value of  $\xi(V)$ indicates a strong deviation from this assumption. Given $\xi(V)$, $\beta_0(V)$ remains identifiable even without Assumption~\ref{ass:equiconf}, using a formula similar to \eqref{eqn:beta0}, except that the numerator of the first term is modified to $\beta_1(V)\sigma^2_1(V) - \xi(V)$ to account for the violation.
Further details on  sensitivity analysis are provided in Section S2 of the supplementary material. 

	\section{Multiply robust estimation}\label{sec:mr}
	
    {In this section, we focus on the statistical estimands $\tau^\ho$ and $\tau^\eq$, which can be interpreted as the ATE in the target population under the identification assumptions.}    
    While the nuisance models in Lemma \ref{lem:homo} and Theorem \ref{thm:identification} can be estimated nonparametrically using methods like kernel smoothing or series estimation, these approaches may suffer from poor performance in moderate- or high-dimensional settings due to the curse of dimensionality.
To construct estimators with better finite-sample properties, we turn to semiparametric efficiency theory for estimation. Any regular and asymptotically linear estimator has an influence function, and the one with the lowest variance is known as the efficient influence function (EIF) \citep{bickel1993efficient}. The EIF is central to the semiparametric theory, as its variance defines the semiparametric efficiency bound. It also provides the basis for constructing locally efficient, multiply robust (MR) estimators, which maintain consistency even when some components of the observed data distribution are misspecified, making them particularly useful in practice.

{We consider the  nonparametric observed data  model
$    \calM=\{f(O): \text{$f(O)$ is unrestricted}\}$.
We refer to $\calM$ as the  nonparametric model because no parametric models are imposed in $\calM$ except some positivity and support overlap assumptions.
In the following subsections, we derive the EIFs for the statistical estimands $\tau^\ho$ and $\tau^\eq$ under 
$\calM$, relying only on standard conditions such as positivity and support overlap in the observed data distribution. These conditions do not restrict the model structure and are distinct from the stronger assumptions imposed at the identification stage.}

	\subsection{MR estimation of $\tau^\ho$}\label{subsec:mr-homo}

In this subsection, {we develop the MR estimator of  $\tau^\ho$ expressed in \eqref{eqn:identify-homo}.}
The estimator relies on the results in \cite{wang2018bounded}. Therefore, this section effectively serves as a review of their MR estimator,
 adapted to our  data fusion setting. 
 {We first derive the EIF of $\tau^\ho$  in $\calM$,}
    and let $\Delta^\ho=(\pi,\mu_0^X,\mu_0^Y,\delta^X,\beta_{1},\omega)$ denote the set of all nuisance functions associated with this EIF. {Among these, the two conditional marginal distributions  $Y \mid Z, V, R = 1$ and $X \mid Z, V, R = 1$ are not derived from a natural factorization of the joint distribution, although they still remain variationally independent by Sklar’s theorem \citep{sklar1959fonctions}.}
    
    \begin{lemma}
    \label{lem:eif-homo}
       The EIF for  {$\tau^\ho$ 
       in $\calM$
       is given by $\psi_{\eif}(O;\Delta^\ho,q)-q^{-1}(1-R)\tau^\ho$,} where 
            \begin{align*}
		\psi_{\eif}(O;\Delta^\ho,q)
			=\frac{1-R}{q}\beta_1(V)+\frac{\gamma(V)}{q}\psi_{1}(O;\Delta^\ho),    
	\end{align*}
    and
    \begin{align*}			\psi_1(O;\Delta^\ho)=\frac{R(2Z-1)}{f(Z\mid V,R=1)} \frac{1}{\delta^X(V)}\big\{Y-\mu_0^Y(V)- X\beta_1(V)+\mu_0^X(V)\beta_1(V) \big\}.
		\end{align*}
    \end{lemma}

{The EIF in Lemma \ref{lem:eif-homo} shares a similar structure with Theorem 2 of \cite{li2023efficient}. Under the additional requirement that the  target distribution aligns with the observed distributions from both data sources in a specific manner, our result can be viewed as a special case when combined with the EIF derived in \cite{wang2018bounded}.  However, our formulation does not require this alignment. For completeness, we provide a direct and self-contained  proof of Lemma \ref{lem:eif-homo} in the supplementary material.}
The EIF must have mean zero and therefore, Lemma \ref{lem:eif-homo} implies that
$\tau^\ho = E\{\psi_\eif(O;\Delta^h,q)\}$.
    Let $\Pn$ denote the empirical average, for example,
	$\Pn \{g(O)\} = n^{-1}\sum_{i=1}^n g(O_i)$ for any function $g(\cdot)$.
     When nuisance functions in $\Delta^\ho$ have been estimated by $\hat\Delta^\ho$ and $\hat q=n^{-1}\sum_{i=1}^n(1-R_i)$,
	 we can construct a MR estimator of $\tau^\ho$  as 
    \begin{align}\label{eqn:mr-homo}
	\hat\tau_{\mr}^\ho=	\Pn \{\psi_{\eif}(O;\hat\Delta^\ho,\hat q)\}.
	\end{align}
  The strategy of constructing estimators from estimating equations based on influence functions has been widely used in  causal inference literature  \citep{wang2018bounded,shi2020multiply,sun2022selective,sun2022semiparametric,jiang2022multiply}. 
  
  Following \cite{wang2018bounded}, we apply a parametric model-based approach to estimate the nuisance functions in $\Delta^\ho$. Specifically, we consider parametric  models $\Delta^\ho(\Theta_1)$ with $\Theta_1=(\theta_1^\T,\ldots,\theta_{6}^\T)^\T$, where $\theta_j$ is a vector of finite-dimensional parameters associated with the $j$th nuisance model in $\Delta^\ho$ for $j=1,\ldots,6$; that is, 
	$\Delta^\ho(\Theta_1)=\{\pi(v;\theta_1),\mu_0^X(v;\theta_2),\mu_0^Y(v;\theta_3),\delta^X(v;\theta_4),\beta_1(v;\theta_5),\omega(v;\theta_6)\}.$ We first use maximum likelihood estimation to obtain estimators $\hat\theta_1,\hat\theta_2,\hat\theta_3,\hat\theta_6$ of the parameters in the nuisance models $\pi,\mu_0^X,\mu_0^Y,\omega$, because these nuisances are either models for binary variables or simply have  conditional mean forms.
Next, similar to \cite{wang2018bounded}, we  solve the following estimating equations  to obtain estimators $\hat\theta_4$ and $\hat\theta_5$ of the  parameters associated with  models $\delta^X$ and $\beta_1$:
	{\small \begin{align}
		&\Pn \bigg[ D_4(V) \Big\{X - \delta^X(V;\theta_4)Z-\mu_0^X(V;\hat\theta_2)\Big\}\frac{R(2Z-1)}{f(Z\mid V,R=1;\hat\theta_1)}\bigg]=0, \label{eqn:theta3} \\ 
		&\Pn \bigg[  D_5(V) \Big\{Y-\beta_1(V;\theta_5) X-\mu_0^Y(V;\hat\theta_3)+\mu_0^X(V;\hat\theta_2)\beta_1(V;\theta_5)\Big\}\frac{R(2Z-1)}{f(Z\mid V,R=1;\hat\theta_1)}\bigg]=0, \label{eqn:theta5}
	\end{align}}
	where $D_j(V)$ is a user-specified vector function of the same dimension as $\theta_j$ for $j=4,5$.  Let $\hat\Theta_1 = (\hat\theta_1^\T,\ldots,\hat\theta_{6}^\T)^\T$ and $\hat\Delta^\ho=\Delta^\ho(\hat\Theta_1)$.
   Then the MR estimator $\hat\tau_{\mr}^\ho$ under homogeneous CATE is  given by \eqref{eqn:mr-homo}.

	\begin{lemma}\label{lem:mr-homo}
        The estimator		$\hat\tau_{\mr}^\ho$ is consistent and asymptotically normal  under standard regularity conditions if one of the following  holds:  

        \begin{itemize}
            \item[(i)] $\calM_{1}^\ho$: models for   $(\pi,\delta^X,\omega)$ are correct; 

       \item[(ii)]    $\calM_{2}^\ho$: models for $(\mu_0^X,\mu_0^Y,\beta_1)$ are correct;

        \item[(iii)] $\calM_{3}^\ho$:  models for $(\pi,\beta_1)$ are correct.
        \end{itemize}
        



            \noindent
	In addition, if all the conditions in (i)--(iii) hold,  then $\hat\tau_{\mr}^\ho$ attains the semiparametric efficiency bound. 
	\end{lemma}

  Like \cite{wang2018bounded}, which considered a single dataset with IVs, Lemma \ref{lem:mr-homo} shows that we also obtain a triply robust estimator in the current data fusion setting under the homogeneous CATE assumption. This estimator ensures consistency when one of the three model sets is correctly specified. However, since our focus is on data fusion,  our first model set $\calM_1^\ho$ includes an additional model for the data source mechanism $\omega$. The other two model sets are similar to those in \cite{wang2018bounded}.
    

    \subsection{MR estimation of $\tau^\eq$}\label{subsec:mr-equi}
    In this subsection, {we develop the MR estimator of $\tau^\eq$ expressed in \eqref{eqn:identify-equi}.}
    Similar to Section \ref{subsec:mr-homo}, we first derive
   the EIF for {$\tau^\eq$   
    in
    $\calM$.}
	Let $\Delta^\eq=(\Delta^\ho,\mu_0,\beta_{0},\phi_0, \rho)$ denote the set of all nuisance functions associated with this EIF. 
    The nuisance functions in $\Delta^\eq$ are variationally independent.
	Using the nuisances in $\Delta^\eq$, we can express the functions $\mu_1(v)$ and $\phi_1(v)$ as
	\begin{align}\label{eqn:mu-phi}
		\mu_1(v)=\delta^X(v)\pi(v)+\mu_0^X(v), \quad 
        \phi_1(v)=\mu_0^Y(v)-\beta_1(v)\mu_0^X(v).	
	\end{align}

	\begin{theorem}
    \label{thm:eif}
		The EIF for  {$\tau^\eq$  
        in $\calM$
        is given by $\varphi_{\eif}(O;\Delta^\eq,q)-q^{-1}(1-R)\tau^\eq$,} where
		\begin{align*}
			\varphi_{\eif}(O;\Delta^\eq,q)
			=&
			\frac{1-R}{q}\beta_0(V)-\bigg\{\frac{R}{f(R\mid V)}-1 \bigg\}\frac{\eta \{Y-\beta_R(V)X-\phi_R(V)\}-\rho(V)}{q\sigma^2_0(V)}
			\\
			&
			+ \frac{\gamma(V)\sigma^2_1(V)}{q\sigma^2_0(V)}\psi_{1}(O;\Delta^\ho),
 		\end{align*}
        and $\psi_1(O,\Delta^\ho)$ is defined in Lemma \ref{lem:eif-homo}.
	\end{theorem}

	The function $\psi_1(O;\Delta^\ho)$ involves only the auxiliary sample and also appears in the EIF for $\tau^\ho$, as shown in Lemma \ref{lem:eif-homo}. Compared with Lemma \ref{lem:eif-homo},
	the expression of the EIF for $\tau^\eq$ presented  here is more complicated.  Under homogeneous CATE with $\beta_0(V)=\beta_1(V)$, the first term of the EIF in Theorem \ref{thm:eif} coincides with that
     in Lemma \ref{lem:eif-homo}. However, because we do not impose the homogeneous CATE assumption when identifying $\tau$ via \eqref{eqn:identify-equi},  the corresponding EIF includes additional functions, as shown in the second  term of $\varphi_{\eif}(O;\Delta^\eq,q)$.
    Define
	the following semiparametric models that impose certain restrictions on the observed data
	distribution:
    \begin{itemize}
    
    \item[$\calM_{1}^\eq$]: models 
    for $(\pi,\mu_0^X,\delta^X)$
    and $(\omega,\mu_0)$
    are correct;
    
   \item[$\calM_{2}^\eq$]: models 
    for $(\mu_0^X,\mu_0^Y,\beta_1)$
    and $(\omega,\mu_0)$
    are correct;

	\item[$\calM_{3}^\eq$]: models 
    for 
    $(\pi,\mu_0^X,\delta^X)$
    and $(\omega,\beta_0,\phi_0)$
    are correct;
    
	\item[$\calM_{4}^\eq$]: models 
    for $(\mu_0^X,\mu_0^Y,\beta_1)$
    and $(\omega,\beta_0,\phi_0)$
    are correct;

	\item[$\calM_5^\eq$]: models for 
    $(\pi,\mu_0^X,\delta^X)$
    and $(\mu_0,\beta_0, \rho)$
    correct;	
    
	\item[$\calM_6^\eq$]: models for 
$(\mu_0^X,\mu_0^Y,\beta_1)$
    and $(\mu_0,\beta_0, \rho)$
    correct.
    
	\end{itemize}
	%
	\noindent
Let \( \calM_{\union}^\eq = \cup_{k=1}^6 \calM_k^\eq \) denote the semiparametric model that holds if any of \( \{\calM_k^\eq\}_{k=1}^6 \) is true. We can show that if \( \cup_{k\neq 5} \calM_k^\eq \) holds, then \( \cup_{k=1}^3\calM_k^\ho \), as defined in Lemma \ref{lem:mr-homo}, also holds. However, if only \( \calM_5^\eq \) holds, then \( \cup_{k=1}^3\calM_k^\ho \) may not 
be true.
	Similar to Section \ref{subsec:mr-homo}, we  consider modeling 
    $\Delta^\eq$ using standard parametric models
	$\Delta^\eq(\Theta)$ with $\Theta=(\Theta_1^\T, \theta_7^\T,\theta_8^\T,\theta_9^\T,\theta_{10}^\T)^\T$, where $\Theta_1$ is defined in Section \ref{subsec:mr-homo} and $\theta_j$ is a vector of finite-dimensional parameters associated with the $j$th nuisance model in $\Delta^\eq$ for $j=7,\ldots,10$; that is, 
	$\Delta^\eq(\Theta)=\{\Delta^\ho(\Theta_1), \mu_0(v; \theta_7),$\\$
	\beta_0(v;\theta_8), \phi_0(v;\theta_9),
	\rho(v;\theta_{10})\},$ with  $\Delta^\ho(\cdot)$  defined in Section \ref{subsec:mr-homo}.
	
    Below we develop a three-step procedure	to estimate the nuisance parameters.	
	In the first step, besides the estimators $\hat\theta_1,\hat\theta_2,\hat\theta_3,\hat\theta_6$ obtained in Section \ref{subsec:mr-homo}, we also  use maximum likelihood estimation to obtain  the estimator $\hat\theta_7$ of the parameter  in the nuisance model $\mu_0$.	
	In the second step, we   obtain  estimators $\hat\theta_4$ and $\hat\theta_5$ of the  parameters associated with  models $\delta^X$ and $\beta_1$ by solving \eqref{eqn:theta3} and \eqref{eqn:theta5}, respectively.
	We use $\hat\pi(v)$ to simply denote $\pi(v;\hat\theta_1)$, and other estimated nuisance models are denoted similarly.
	After obtaining the  estimators of the first seven nuisance functions in $\Delta^\eq$, we then plug them into \eqref{eqn:mu-phi} to  obtain the estimators $(\hat\mu_1,\hat\phi_1)$ of $(\mu_1,\phi_1)$. Similarly, we can obtain the estimators $(\hat\eta,\hat\sigma^2_1,\hat\gamma)$ by plugging the estimated nuisance functions into the corresponding expressions of $(\eta,\sigma_1^2,\gamma)$ defined in Table \ref{tab:notation}; that is, 
	\begin{align*}
		&\hat\eta=X-R\hat\mu_1(V)-(1-R)\hat\mu_0(V),
		~~ \hat\sigma^2_1(V)=\hat\mu_1(V)\{1-\hat\mu_1(V)\},~~\hat\gamma(V)=\frac{1-\hat\omega(V)}{\hat\omega(V)}.
	\end{align*}
	Then we  obtain a plug-in estimator of $\psi_1(O;\Delta^\ho)$ defined in Lemma~\ref{lem:eif-homo} and denote it by $\hat\psi_1$. 
	
	In the third step, we  estimate the remaining  nuisance models $(\beta_0,\phi_0,\rho)$. We first define  two functions based on  estimates of some nuisance models:
	\begin{align*}
		&\hat\beta(\theta_8)=R\hat\beta_1(V)+(1-R)\beta_0(V;\theta_8),\qquad \hat\phi(\theta_9)=R\hat\phi_1(V)+(1-R)\phi_0(V;\theta_9).
	\end{align*}	
	Then we 	solve the following equations in \eqref{eqn:theta8}--\eqref{eqn:theta10} to obtain estimators $\hat\theta_8$, $\hat\theta_9$, and $\hat\theta_{10}$ of the parameters associated with the three nuisance models $\beta_0$, $\phi_0$, and $\rho$:
	{\small	
		\begin{align}
			&\Pn \bigg[ D_8(V) \Big\{\Big(\frac{R}{\hat f(R\mid V)}-1 \Big)\big(\hat\eta (Y-\hat\beta(\theta_8)X-\hat\phi(\theta_9))
			-\rho(V;\theta_{10}) \big)-\hat\gamma(V)\hat\sigma^2_1(V)\hat\psi_1\Big\}\bigg]=0,\label{eqn:theta8}\\
			&\Pn \Big[  D_9(V) (1-R)\big\{Y-\beta_0(V;\theta_8)X-\phi_0(V;\theta_9)\big\}\Big]=0, \label{eqn:theta9}\\
			&\Pn\Big[D_{10}(V)(1-R)\big\{(X-\hat\mu_0(V))(Y-\beta_0(V;\theta_8)X)-\rho(V;\theta_{10}) \big\} \Big]=0,&\label{eqn:theta10}
		\end{align}
	}
	where $\hat f(R\mid V)=R\hat\omega(V)+(1-R)\{1-\hat\omega(V)\}$, and
	$D_j(V)$ is a user-specified vector function of the same dimension as $\theta_j$ for $j=8,9,10$. 
    The following lemma shows consistency of the proposed estimators for the nuisance parameters under certain conditions.
	
	\begin{lemma}\label{lem:consistencynuisance}
		Under standard regularity conditions, the estimators obtained from the three-step procedures for the parameters in $\calM^\eq_j$ are consistent if $\calM^\eq_j$ holds, for $j=1,\ldots,6$.	
	\end{lemma}
	Let $\hat\Theta = (\hat\Theta_1^\T,\hat\theta_7^\T,\hat\theta_8^\T,\hat\theta_9^\T,\hat\theta_{10}^\T)^\T$ and $\hat\Delta^\eq=\Delta^\eq(\hat\Theta)$.
    Since the EIF has mean zero, Theorem \ref{thm:eif} implies that $\tau^\eq=E\{\varphi_{\eif}(O;\Delta^\eq, q)\}$.
	Accordingly, the corresponding MR estimator of $\tau^\eq$  is  given by
	$\hat\tau_{\mr}^\eq=	\Pn\big\{\varphi_{\eif}(O;\hat\Delta^\eq,\hat q)\big\}$.
	\begin{theorem}\label{thm:mr}
        The estimator
		$\hat\tau_{\mr}^\eq$ is a consistent and asymptotically normal estimator of $\tau^\eq$ in $\calM^\eq_{\union}= \cup_{k=1}^6 \calM_k^\eq $ under standard regularity conditions.
	In addition, if all the models in $\Delta^\eq$  are correct, then $\hat\tau_{\mr}^\eq$ attains the semiparametric efficiency bound.
	\end{theorem}

	The estimator $\hat\tau_\mr^\eq$ offers six genuine opportunities to estimate $\tau^\eq$ consistently
	and  also has valid inference provided that one of the six strategies holds. 
    Under particular specifications of the nuisance models, $\hat\tau_{\mr}^\eq$ reduces to other different semiparametric estimators. 	
	There are six such estimators $\hat\tau_1^\eq,\ldots,\hat\tau_6^\eq$, dependent on specific model specifications in $\calM^\eq_1,\ldots,\calM^\eq_6$, respectively. Under standard regularity conditions, $\hat\tau_j^\eq$ is consistent and asymptotically normal in $\calM^\eq_j$ for $j=1,\ldots,6$. More details about these semiparametric estimators can be found in Section S1 of the supplementary material.
	

	\subsection{Flexible MR estimation
    }\label{sec:flexible-mr}
    
	An important advantage of  MR estimators constructed using
	the EIF is that they often yield second-order bias terms, which allow for slow convergence
	rates for the nuisance parameters involved, thereby enabling the use of flexible regression
	techniques in estimating these quantities \citep{chernozhukov2018double,kennedy2022semiparametric}.	In this section, we introduce the flexible MR estimation  about $\tau^\eq$
    using machine learning methods. A similar estimation approach  applies to the other estimand $\tau^\ho$.
    

To construct our estimator, we use cross-fitting for nuisance estimation. Let \( K \) be a fixed positive integer. We divide the data into \( K \) groups, train nuisance models on all but one fold, and apply the estimated nuisances only to the hold-out fold.
Specifically, we create \( K \)-fold random partitions for the auxiliary data \(\{i: R_i = 1\}\) and the target data \(\{i: R_i = 0\}\), with index sets \(\{\mathcal{I}_k^1\}_{k=1}^K\) and \(\{\mathcal{I}_k^0\}_{k=1}^K\), respectively. The combined partition \(\{\mathcal{I}_k = \mathcal{I}_k^1 \cup \mathcal{I}_k^0\}_{k=1}^K\) divides the entire sample \(\{1, \dots, n\}\) into subsets of sizes \(\{n_k\}_{k=1}^K\), and we define the complement set \(\mathcal{I}_{-k} = \{1, \dots, n\} \setminus \mathcal{I}_k\) with size \( n_{-k} \). The empirical means in the \( k \)th partition and the remaining sample are denoted by \( \mathbb{P}_{n_k} \) and \( \mathbb{P}_{n_{-k}} \), respectively, and we define \( \hat{q}_{-k} = \mathbb{P}_{n_{-k}}(1 - R) \).

For each $k$, we estimate the nuisance functions using the data with index set \(\mathcal{I}_{-k}\). Let \( \hat{\Delta}_{-k}^\eq \) denote the machine learning estimators for nuisance models. First, we obtain estimators \( (\hat{\pi}_{-k}, \hat{\mu}_{0,-k}^X, \hat{\mu}_{0,-k}^Y, \hat{\omega}_{-k}, \hat{\mu}_{0,-k}) \) for the conditional mean functions \( (\pi, \mu_0^X, \mu_0^Y, \omega, \mu_0) \).
Next, given \( (\hat{\pi}_{-k}, \hat{\mu}_{0,-k}^X, \hat{\mu}_{0,-k}^Y) \), we estimate \( {\delta}^X \) and \( {\beta}_1 \) using machine learning methods based on the following conditional mean relationships:
\begin{align*}
    \delta^X(V) = E\left\{\frac{X - \mu_0^X(V)}{\pi(V)} \mid V, R=1 \right\}, \quad
    \beta_1(V) = E\left\{\frac{Y - \mu_0^Y(V)}{\pi(V) \delta^X(V)} \mid V, R=1 \right\}.
\end{align*}
Then, using the estimators \( (\hat{\pi}_{-k}, \hat{\mu}_{0,-k}^X, \hat{\omega}_{-k}, \hat{\mu}_{0,-k}, \hat{\delta}_{-k}^X, \hat{\beta}_{1,-k}) \), we  estimate \( {\beta}_{0} \), \( {\phi}_{0} \), and \( {\rho} \) based on the expression in \eqref{eqn:beta0} and the definitions of \( (\phi_0, \rho) \) in Table \ref{tab:notation}.
Finally, the MR estimator of $\tau^\eq$, using cross-fitting, is given by: 
\begin{align*}
    \hat{\tau}_{\cf}^\eq &= \frac{1}{K} \sum_{k=1}^K \mathbb{P}_{n_k} \big\{
    \varphi_{\text{eff}}(O; \hat{\Delta}_{-k}^\eq, \hat{q}_{-k}) \big\}.
\end{align*}
Let \( \| g \| = \{\int g^2(o) d\mathbb{P}(o) \}^{1/2} \) denote the \( L_2(\mathbb{P}) \) norm of any function \( g(\cdot) \). In the following theorem, we establish the asymptotic properties of the cross-fitting estimator \( \hat{\tau}_{\cf}^\eq \).

\begin{theorem}\label{thm:cross-fitting-asymptotic}
    For each $k$, we  assume the following conditions hold:
	\begin{itemize}
		\item [(i)]
    
        $\hat\Delta_{-k}^\eq(v) \rightarrow \Delta^\eq(v)$ in probability for all $v$ in the support of $V$;
		\item[(ii)] 

        $c<\{\hat\pi_{-k}(v), \pi(v)\}<1-c$, $\{\hat\mu_{0,-k}(v),\mu_0(v), \hat\omega_{-k}(v),\omega(v),|\hat\delta_{-k}^X(v)|,|\delta^X(v)|\}>c$, and \\$\{|\hat\beta_{1,-k}(v)|,|\beta_1(v)|\}$$< C$ for some $c\in (0,1)$, $C>0$, and all $v$ in the support of $V$;
		\item[(iii)] 

        $n^{1/2}$-convergence of second-order terms, i.e.,
		\begin{align*}
			&\big(\Vert\hat\beta_{1,-k}-\beta_1\Vert +\Vert\hat\mu_{0,-k}^Y-\mu_0^Y\Vert +\Vert \hat\mu_{0,-k}^X-\mu_0^X\Vert \big) \big(\Vert\hat\pi_{-k}-\pi\Vert +\Vert\hat\delta_{-k}^X-\delta^X\Vert +\Vert \hat\mu_{0,-k}^X-\mu_0^X\Vert \big)\\
			&~~~ + \Vert\hat\mu_{0,-k}-\mu_0\Vert \big(\Vert\hat\beta_{0,-k}-\beta_0\Vert +\Vert \hat\phi_{0,-k}-\phi_0\Vert \big)+\Vert \hat\omega_{-k}-\omega\Vert \Vert\hat\rho_{-k}-\rho\Vert=o_\mathbb{P}\big(n^{-1/2}\big).
		\end{align*}
	\end{itemize}
	Then $\hat{\tau}_{\cf}^\eq$ is asymptotically normal,
    and achieves the semiparametric efficiency bound.
\end{theorem}

Conditions (i)-(iii) are similar to those for double machine learning estimation of average treatment effects \citep{kennedy2017non,chernozhukov2018double,kennedy2022semiparametric}. These conditions are 
	all common in standard non-parametric regression problems \citep{van2000asymptotic}. Among them,  
	condition (iii) is satisfied if  all
	the nuisance parameters are estimated with root mean-squared error rates diminishing faster
	than $n^{-1/4}$.
	Such rates are  achievable for many highly data-adaptive machine learning methods, including lasso, 	
	neural networks 
	or ensembles of these methods.

	\section{Simulation studies}\label{sec:simulation}
	
	In this section, we conduct simulation studies to evaluate the finite-sample performance of the proposed estimators. In our simulation setting, baseline covariates $V=(V_1,V_2)$ are independently generated from  standard normal distribution $N(0,1)$. The case with four-dimensional covariates is shown in Section S4.2 of the supplementary material.    
    We consider the functional form $V_k^*=\{1+\exp(1-2V_k)\}^{-1}$ for $k=1,2$. The unmeasured confounder $U$ is generated from a uniform distribution  $\unif\{-a(V^*),a(V^*)\}$, where $a^2(V^*)=0.75+1.5V_1^*+1.5V_2^*$.
	Then the sampling indicator $R$, instrumental variable $Z$ in the auxiliary sample, treatment $X$, and outcome $Y$ are separately generated by the following mechanism:
	\begin{align*}
		R\mid V&\sim \text{Bernoulli}\{p=(1+\exp(-0.8 - 0.5V_1^* + 2.5V_2^*))^{-1}\},\\
		Z\mid V,R=1&\sim \text{Bernoulli}\{p=(1+\exp(0.5 +0.5 V_1^* - 3 V_2^*))^{-1}\},\\
		X\mid Z,V,U,R&\sim \text{Bernoulli}\{p=R\alpha_1(Z)+(1-R)\alpha_0(V)+0.1U\},\\
		Y\mid X,V,U,R&\sim N\{(2 - \kappa R + 2V_1^* -0.5 V_2^*) (1 + X) +U, 0.5^2\},
	\end{align*}
	where $\alpha_1(Z)=\{1+\exp(-1.3+2.4Z)\}^{-1}$, $\alpha_0(V)=\{1+\exp(-V_1^* + V_2^*)\}^{-1}$, and $\kappa=\beta_0(V)-\beta_1(V)\in\{0,1,2\}$ denotes the difference of CATEs between two populations. Under the above data generating mechanism, the equi-confounding assumption holds according to Example \ref{exam:model}. 
    We also observe that correct models for the nuisance models $(\pi,\mu_0^X,\omega,\mu_0)$ in $\Delta^\eq$ are given by a logistic regression with $V^*$ as linear predictors, and correct models for  other models $(\mu_0^Y,\delta^X,\beta_1,\beta_0,\phi_0,\rho)$ in $\Delta^\eq$ are given by a linear regression with $V^*$ as predictors. We  are interested in estimating the average causal effect  $\tau$, whose true value is 2.43. We  evaluate the performance of the proposed estimators in situations
	where some models may be misspecified. 
    In the misspecification case, the quadratic functional form $V_k^{**}=(V_k-0.5)^2$ is used  in place of $V_k^*$  for $k=1,2$ when constructing the nuisance models in $\Delta^\eq$. To estimate the nuisance models based on \eqref{eqn:theta3}-\eqref{eqn:theta5} and \eqref{eqn:theta8}-\eqref{eqn:theta10} given in Section \ref{sec:mr}, we set $D_j(V)=(1,V^*)^\T$ if the  model is correct and $D_j(V)=(1,V^{**})^\T$ otherwise.
	

	\begin{table}[t!]
		\caption{Bias, standard deviation (SD), and 95\% coverage probability (CP) of the  estimators under $\kappa=0$ and   $n=2000,4000$. All values have been multiplied by $100$. 
		}
		\label{tab:sim-2V-betadf0}
		\centering
		\begin{threeparttable}    
			\renewcommand{\arraystretch}{0.75}
			\resizebox{0.98\columnwidth}{!}{%
				\begin{tabular}{ccccccccccccccccccccccccc}
					\toprule
					
					
					\multirow{3}{*}{Scenario} & 
					\multicolumn{1}{c}{} &
					\multicolumn{2}{c}{$\hat\tau^\eq_{1}$} & 
					\multicolumn{1}{c}{} &
					\multicolumn{2}{c}{$\hat\tau^\eq_{2}$} & 
					\multicolumn{1}{c}{} &
					\multicolumn{2}{c}{$\hat\tau^\eq_{3}$}& 
					\multicolumn{1}{c}{} &
					\multicolumn{2}{c}{$\hat\tau^\eq_{4}$}& 
					\multicolumn{1}{c}{} &
					\multicolumn{2}{c}{$\hat\tau^\eq_{5}$}& 
					\multicolumn{1}{c}{} &
					\multicolumn{2}{c}{$\hat\tau^\eq_{6}$}& 
					\multicolumn{1}{c}{} &
					\multicolumn{2}{c}{$\hat\tau_{\mr}^\eq$}& 
					\multicolumn{1}{c}{}&
					\multicolumn{2}{c}{$\hat\tau_{\mr}^\ho$}  \\
					\specialrule{0em}{1pt}{1pt}

					\cmidrule(lr){3-4} 	\cmidrule(lr){6-7}	\cmidrule(lr){9-10}	\cmidrule(lr){12-13} \cmidrule(lr){15-16} \cmidrule(lr){18-19} \cmidrule(lr){21-22}
                    \cmidrule(lr){24-25}
					
					\addlinespace[1.5mm]
					& &	  2000  & 4000 	&& 2000  & 4000	&& 2000   & 4000 && 2000   & 4000 && 2000  & 4000	&& 2000   & 4000 && 2000   & 4000 && 2000   & 4000\\
					\specialrule{0em}{0pt}{2pt}
					\midrule
					
					&	&\multicolumn{1}{c}{}&	\multicolumn{21}{c}{Bias}  \\
					$\calS_0$&
					& -2 &-2 & &0 &-1 & &-2 &-2& & 0& -1&   & -1& -1& &0 &0&&1&1 &&1&1\\
                    	$\calS_1 $&
					& -2 &-2 & &72 & 69& &25 &24& &97 &94 &   &3 &3 & &30 &28&&1& 0& &0 &0\\
						$\calS_2 $&
					& -176 &-180 & &0 &-1 & &-138 &-142& &26 &25 &   &30 &28 & &3 &3&&0&-1 & &0 &-1\\
                    	$\calS_3 $&
					& 24 &24 & &97 &94 & &-2 &-2& & 74& 72&   &-13 &-14 & & -35&-37&&1&0 & &0 &0\\
                    	$\calS_4 $&
					& -138 &-143 & & 26&25 & &-178 &-182& &0 &-1 &   &86 &83 & &-13 &-13&&0&-1 & &0 &-1\\
                    	$\calS_5 $&
					& 3 &3 & &21 & 20& &12 &12& &54 & 52&   &-1 &-1 & &75 &73&&0&0 & &14 &14\\
                    	$\calS_6 $&
					& 31 & 28& &3 &2 & &18 &16& &12 &12 &   &-143 &-148 & & 0&0&&0& -1& &0 &0\\
                    	$\calS_7 $&
					&51  & 50& &54 &52 & &52 &50& &54 &52 &   &54 &52 & & 54&53&&53&51 & &48 &47\\
					\addlinespace[1mm]
					&	&\multicolumn{1}{c}{}&	\multicolumn{21}{c}{SD}  \\
					$\calS_0$&
					& 45 &31 & &19 &13 & &45 &31& & 19& 13&   & 36& 24& &19 &13&&24&15 &&25&16\\
					$\calS_1 $&
					& 45 & 31& &25 &17 & &43 &29& &24 &17 &   &20 &14 & &18 &13&&32&20 & & 32&21\\
						$\calS_2 $&
					& 50 &36 & &19 & 13& &46 &33& &19 &13 &   &23 &16 & &14 &10&&19&14 & & 20&14\\
                    	$\calS_3 $&
					& 43 &29 & & 24&17 & &45 &31& &25 &17 &   &20 &13 & &28 &19&&30&20 & &32 &21\\
                    	$\calS_4 $&
					& 46 &33 & &19 &13 & &50 &36& & 19& 13&   &38 &26 & & 18&12&&19& 13& &20 &14\\
                    	$\calS_5 $&
					& 20 &14 & &19 &13 & &20 &14& &19 & 13&   &36 &24 & &26 &18&&25&16 & &16 &11\\
                    	$\calS_6 $&
					& 24 & 17& &14 &10 & &20 &14& &14 &10 &   & 43& 31& &19 &13&&19&13 & &19 &13\\
                    	$\calS_7 $&
					& 19 & 13& &18 &13 & &19 &13& &18 &13 &   & 19&13 & &18 &13&&18&13 & &18 &13\\
					\addlinespace[1mm]
					&	&\multicolumn{1}{c}{}&	\multicolumn{21}{c}{CP}  \\
					$\calS_0 $& 
					& 97 &97 & &97 &96 & &97 &97& & 97& 96&   & 98&98 & &97 &96&&98&96 &&97&96\\
                    $\calS_1 $&
					& 97 & 97& &13 & 1& &96 &92& &0 &0 &   &99 &99 & & 67&36&&95& 95& &94 &95\\
						$\calS_2 $&
					& 2 &0 & &97 &96 & &10 &0& & 73&51 &   &81 &62 & &97 &97&&97&96 & &96 &96\\
                    	$\calS_3 $&
					& 96 &92 & &0 & 0& & 97&97& &10 &0 &   &94 &87 & &80 &53&&95&96 & &94 &95\\
                    	$\calS_4 $&
					& 9 &0 & &72 &52 & &2 &0& &97 & 96&   &35 &5 & & 93&85&&97&96 & &96 &96\\
                    	$\calS_5 $&
					& 99 &99 & & 82&69 & &95 &91& &15 &1 &   &98 &98 & &13 &1&&96&97 & &87 &76\\
                    	$\calS_6 $&
					& 81 &66 & &98 &97 & &90 &85& &89 &79 &   &5 &0 & & 97&96&&97&96 & &96 &96\\
                    	$\calS_7 $&
					& 20 & 2& &13 &1 & &20 &2& & 14& 1&   &13 &0 & & 11&0&&14&1 & &22 &2\\
					\addlinespace[1mm]
					
					\bottomrule
				\end{tabular}
			} 
		\end{threeparttable}       
	\end{table}

     We apply the proposed MR estimators $\hat\tau_{\mr}^\ho$, $\hat\tau_{\mr}^\eq$ introduced in Section \ref{sec:mr} and other semiparametric estimators $\{\hat\tau^\eq_j\}_{j=1}^6$ introduced in Section S1 of the supplementary material
    to estimate $\tau$. We consider the following eight scenarios: 
      
       $\calS_0$, 
       all the models are correct;
         
     $\calS_j$, 
     only the models in $\calM^\eq_j$ are correct for $j=1,\ldots,6$. 
        
       $\calS_7$,  
       none of the models is correct.

        \noindent
	Table~\ref{tab:sim-2V-betadf0} summarizes the simulation results for all estimators under the eight scenarios for $\kappa=0$ and $n=2000,4000$, where the bias, standard deviation, and 95\% coverage probability of these estimators averaged across 1000 replications are reported. As expected, the proposed estimator $\hat\tau_\mr^\eq$ exhibits small bias and has coverage probability close to the nominal level  in $\{\calS_j\}_{j=0}^6$, confirming its multiple robustness property. In contrast, the other six estimators $\{\hat\tau^\eq_j\}_{j=1}^6$ generally show substantial bias in scenarios where their respective models are misspecified. For example, the estimator $\hat\tau^\eq_1$ performs well in scenarios $\{\calS_j\}_{j=0,1}$, while exhibiting notable bias in scenarios $\{\calS_j\}_{j=2,3,4,6}$. We also observe that $\hat\tau^\eq_1$ has comparable performance in some  scenarios  where models are misspecified, e.g., $\calS_5$,  to those in scenarios where models are correct. Similar phenomenon may appear in specific scenarios for other estimators, which is subject to the simulation setting. In general, these six estimators can  maintain their consistency only when their corresponding models are correct. 
Since \(\kappa = 0\), the homogeneous CATE assumption holds. We observe that the other MR estimator \(\hat\tau_\mr^\ho\) performs similarly to \(\hat\tau_\mr^\eq\) in scenarios \(\{\calS_j\}_{j \neq 5}\). This is because  in these scenarios, at least one of the conditions \(\{\calM_j^\ho\}_{j=1}^3\) is true. As a result, \(\hat\tau_\mr^\ho\) is consistent according to Lemma \ref{lem:mr-homo}.
However, it shows significant bias and low coverage probability in scenario \(\calS_5\), because none of the conditions \(\{\calM_j^\ho\}_{j=1}^3\) holds in this scenario.
When $\kappa\in\{1,2\}$,
the homogeneous CATE assumption does not hold. The simulation results for \(\hat\tau_{\mr}^\eq\) and \(\{\hat\tau^\eq_j\}_{j=1}^6\) in these two cases are similar to those when \(\kappa = 0\). However, \(\hat\tau_\mr^\ho\) exhibits significant biases in all scenarios of these two cases.  Due to space constraints, the results for \(\kappa \in \{1, 2\}\) are provided in Section S4.1 of the supplementary material.


	We also implement the proposed MR estimator by using machine learning methods for nuisance model estimation. Specifically, we consider estimators $\hat\tau_{\text{cf-la}}^\eq$  and $\hat\tau_{\text{cf-sl}}^\eq$, whereby the nuisance models are estimated by lasso 
	and super learner,
	respectively. Figure~\ref{fig:mrml} shows the corresponding results about mean squared error (MSE) and 95\% coverage probability of these two MR machine learning-based estimators, averaged across 1000 repeated experiments with sample size from 2000 to 20000. We observe that as sample size increases, the MSEs of both estimators become smaller, and the 95\% coverage probabilities are more close to the nominal level.
	These results corroborate our previous theoretical findings and demonstrate the advantages of the proposed MR estimators. 
    A simulation study for assessing the sensitivity analysis method
    proposed in Section S2
    is given in 
    Section S4.3 of 
    the supplementary material.

    \begin{figure}[t!]
		\centering
		\includegraphics[width=\textwidth]{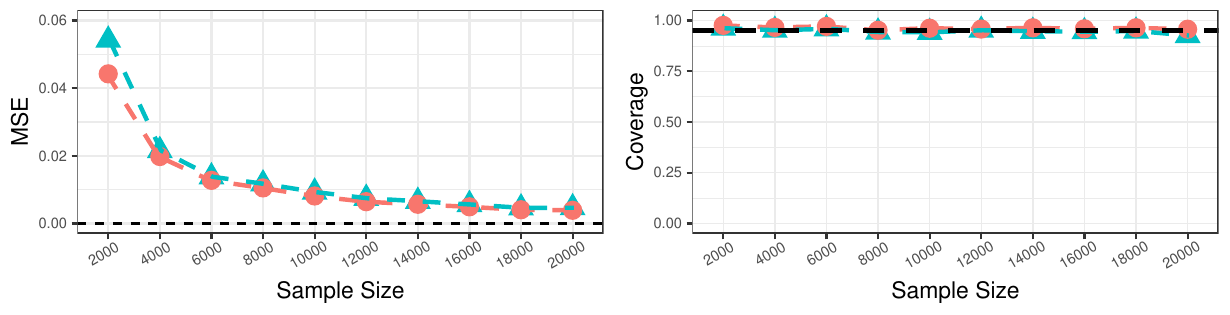}
		\vspace{-1.5cm}
		\caption{
			Mean squared error (MSE) and 95\% coverage probability of the lasso-based MR  estimator $\hat\tau_{\text{cf-la}}^\eq$ (red circle)  and the super learner-based MR estimator  $\hat\tau_{\text{cf-sl}}^\eq$ (green triangle) under $\kappa=0$ and different sample sizes. 			
		}\label{fig:mrml}
	\end{figure}

\section{Application}\label{sec:app}

	In this section, we provide an application of  the proposed approaches by estimating the average causal effect of smoking on physical functional status,
	based on data from the household survey portion of Community Tracking Study  for the years 1996--1997.
    This study is designed to provide information of health care system  change and its effects on people in the United States \citep{kemper1996design}. Smoking is a risk factor of many diseases including heart disease, stroke, diabetes, and most cancers. These diseases affect physical functioning or physical disability. To mitigate possible unmeasured confounding bias, \citet{leigh2004instrumental} used  cigarette price as an IV to assess  the impact of smoking on physical functional status. We found that cigarette price is a strong IV only for the lower-income group but is a  weak IV for the higher-income group. The higher-income group  is defined as those whose incomes are in the top 20\%, with the remaining classified as the lower-income group.
	Below we  use  the data from the lower-income group with strong IV to help estimate the effect of smoking on physical functional status in the higher-income group.

	
	Let the treatment variable be
	$X=1$
	if the subject smokes every day or on some days, and 
	$X=0$ if the subject does not smoke at all.
	The outcome $Y$ is physical functional status, measured by the Physical Component Summary of the SF-12 \citep{ware1995sf}. The baseline covariates include age, gender, race, mental functional status, education, and health condition. However, there may still exist unmeasured variables confounding the relationship between smoking and physical functional status, such as genetic factors. 
    Let the  IV be  $Z=1$ if the cigarette price   is greater than the median and $Z=0$ otherwise. Since our purpose is to estimate the average causal effect among  higher-income individuals, we treat this group  as the target population.
	The target dataset consists of 4198 subjects, with 1123  having $X=1$ and 3075  having $X=0$. The auxiliary dataset includes 18603 lower-income subjects, with 7834 having $X=1$ and 10769  having $X=0$. While the IV (i.e., cigarette price) is available in the higher-income group, it shows weak correlation with the treatment. This is because smoking decisions among higher-income individuals  may not be significantly influenced by cigarette price.
	 In fact, after conducting a logistic regression of the treatment on the IV in the target data and adjusting for observed covariates, we found that the coefficient of the IV is not statistically significant.

    We first apply the classical IV approach \citep{wang2018bounded} to estimate the effect of smoking on physical functional status using the weak IV available in the target data. As discussed in Section~\ref{subsec:preliminary}, the validity of this approach relies on the no-interaction Assumption~\ref{ass:no-interact}, which posits that unmeasured confounders affect the outcome but do not modify the effect of treatment. This assumption appears plausible in our context.
Genetic factors, as noted in prior work (e.g., \cite{fisher1958cancer}), may confound the relationship between smoking and physical function. However, the biological mechanisms through which smoking impairs physical function such as reduced lung capacity operate relatively independently of genetic traits. Empirical literature also suggests that interaction effects in natural systems are typically smaller and less prevalent than main effects.
Furthermore, our analysis adjusts for key effect modifiers, including age, gender, race, education, mental status, and health condition, which are major determinants of both smoking behavior and physical function. By accounting for these variables, the likelihood that remaining unmeasured confounders substantially modify the treatment effect is reduced. Together, these considerations support the plausibility of the no-interaction Assumption \ref{ass:no-interact} in this setting.
	The point estimate  using the weak IV, denoted as $\hat\tau_{\text{weak}}$, is $-9.54$ with a large standard error of 15.69. The corresponding confidence interval $(-40.30, 21.22)$ is very wide.
     This suggests that the weak IV approach may not yield a reliable estimate due to high variability.     
     Fortunately,  the IV is  strongly correlated with the treatment in the lower-income group. The IV estimate for this group is $-10.90$, with a standard error of 3.55, leading to a confidence interval of $(-17.86, -3.94)$. These results suggest that smoking significantly decreases physical functional status in the lower-income population.
     The strong IV information from the lower-income group can therefore be leveraged below to help transport  the causal effect of smoking to the higher-income group.

\begin{table*}[t]
		\caption{Estimates  of the average causal effect $\tau$
        in the higher-income group. \\
		}
		\label{tab: application2-res}
		\centering
        \resizebox{0.9\textwidth}{!}{
			\tabcolsep=0pt
			\begin{tabular*}{\textwidth}{@{\extracolsep{\fill}}lccc@{\extracolsep{\fill}}}
				\toprule
				Estimators      &    Point estimate    & Standard error   & 95$\%$ Confidence interval  \\ 
				\hline
                $ \hat\tau_{\text{weak}}$ & -9.54 & 15.69   & (-40.30,~21.22) \\
                    $ \hat\tau_{\text{mr}}^\ho$ & -7.23 & 3.98 & (-15.03,~0.56) \\
    $ \hat\tau_{\text{mr}}^\eq$ & -8.68 & 4.49  & (-17.48,~0.12) \\
    $ \hat\tau_{\text{cf-la}}^\eq$ & -8.32 & 4.73  & (-17.60,~0.96) \\
    $ \hat\tau_{\text{cf-sl}}^\eq$ & -8.40  & 4.53  & (-17.28,~0.48) \\

				\bottomrule
			\end{tabular*}
		}
	\end{table*}

 Next, we apply the two proposed transportability approaches. As a baseline, we first use the MR estimator $\hat\tau_{\mr}^\ho$ under the homogeneous CATE assumption. The point estimate  is $-7.23$, with a standard error of 3.98 and a confidence interval of $(-15.03,0.56)$.
Under the equi-confounding assumption, we then implement the MR estimator $\hat\tau_{\text{mr}}^\eq$ along with two cross-fitting based estimators, $\hat{\tau}_{\text{cf-la}}^\eq$ and $\hat{\tau}_{\text{cf-sl}}^\eq$, which were also used in the simulation studies. Table \ref{tab: application2-res} summarizes the results for all estimators of the average causal effect $\tau$ in the higher-income group.
The three estimators based on the equi-confounding assumption, $\hat\tau_{\text{mr}}^\eq$, $\hat{\tau}_{\text{cf-la}}^\eq$, and $\hat{\tau}_{\text{cf-sl}}^\eq$, are close to each other. The validity of these estimators depends on  Assumption \ref{ass:equiconf}, which appears reasonable in this application. First, possible unmeasured confounders such as genetic predisposition to smoking, are likely to influence both smoking behavior and physical function similarly across income groups. These factors are not expected to vary systematically between higher- and lower-income populations in a way that would change their confounding effect. Second, our analysis adjusts for important observed covariates that capture major socioeconomic and health-related differences. By accounting for these variables, the residual confounding bias is more likely to remain comparable across groups. We also conduct a sensitivity analysis for the equi-confounding assumption, with results provided in Section S4.4 of the supplementary material.
 
Now we compare and interpret the estimates from the above three approaches, including the weak IV approach and  two transportability approaches. Table \ref{tab: application2-res} shows that the equi-confounding based estimates align in direction with the weak IV estimate but have much smaller standard errors. Although the weak IV estimate is noisy, it serves as an unbiased benchmark. The close alignment between the equi-confounding based estimates and the weak IV result suggests that the transportability of unmeasured bias may be reasonable. Additionally,
the homogeneous CATE assumption yields a similar effect estimate with the equi-confounding based estimators. The two transportability assumptions do not yield conflicting results in this example.
The coherence of results across both approaches enhances confidence in the robustness of our conclusions.
In summary, all point estimates are negative, and the upper bounds of the confidence intervals from all transportability estimators are  slightly above zero. These findings indicate that smoking likely reduces physical functional status in higher-income individuals, with the effect approaching statistical significance.

	\section{Discussion}\label{sec:discussion}
	
	Instrumental variables are widely used to  identify causal effects in the presence of unmeasured confounding. However, the stringent constraints for a valid IV may render such variables unavailable in the target population.  This paper
	proposes two complementary approaches, referred to as the homogeneous CATE approach and the equi-confounding approach, to identify and estimate the average causal effect in the target population using IVs  from an auxiliary population.  For each approach, we have developed a semiparametric efficient multiply robust estimator and investigated its asymptotic properties, even when 
	 highly data-adaptive machine learning techniques are used  for nuisance model estimation. 	

	This paper can be extended or improved in several directions. First, our current analysis focuses on the canonical setting where both the instrumental  and  treatment  variables are binary. Extending our framework to accommodate continuous instrumental and treatment variables poses significant challenges in terms of efficient estimation. In Section S5 of the supplementary material, we have developed a simplified estimation method under linear structural equation models.  
Second, the proposed framework can be extended to incorporate multiple auxiliary datasets with IVs, which may further enhance the efficiency of the proposed estimators.
Finally, although both the homogeneous CATE and equi-confounding assumptions are introduced  for transporting causal effects, either assumption alone is sufficient to ensure identification. If  both assumptions are assumed to hold simultaneously, additional constraints are placed on the observed data distribution. Under such constraints, one might explore more efficient estimation strategies within a semiparametric restricted model.
 We plan to explore these and other related topics in future work.

%

	\section*{Supplementary material}
	
	The supplementary material includes additional semiparametric estimators under equi-confounding, proofs of all theoretical results, additional simulation studies,  sensitivity analysis  with implementations in both simulation and real data application,  discussions on the case of continuous instrumental and treatment variables, and R code for reproducibility.


	\bibliographystyle{apalike}
	
	\bibliography{Bibliography-MM-MC}
\end{document}